# A Gaps Approach to Access the Efficiency and Effectiveness of IT-Initiatives In Rural Areas: case study of Samalta, a village in the central Himalayan Region of India

Kamal Kumar Ghanshala
Chairman,Graphic Era University
Dehradun, Uttarakhand, India

Durgesh Pant
Director, School of
ComputerScience & IT,
Uttarakhand Open University,
Haldwani, Uttarakhand, India

Jatin Pandey
J.R.F., Graphic Era University,
Dehradun, Uttarakhand, India

*Abstract:-***This paper focuses on the effectiveness and efficiency of IT initiatives in rural areas where topology creates isolation to developmental activities. A village is selected for the study and information is gathered through interviews of village dwellers. These collected responses are then analyzed and a gaps model is proposed.**

## I. INTRODUCTION

It is now an all accepted fact that computers have the potential of changing the world in a big way. The latter half of the previous century i.e. 20$^{th}$ century saw an upsurge in the adoption of computing technologies. The countries which embarked on the computing technologies and applications forged ahead in bringing about efficiency and effectiveness in their developmental programmes. All around the world what we witness today are technological interventions. The impact has been so overpowering that the world we are living in has become a techno-world. Information has become the most powerful tool and therefore, Information Technology has acquired unprecedented dimensions. We need technology to get the information as per our needs and requirement.

Over the years, it has blended seamlessly into our psyche, and there are hundreds of tasks that we do every day but do not think about it likewise making a simple phone call, answering an email, video chatting with someone from across the globe, paying bills, automating tasks, and finding information [1].

IT initiatives play an important role in formulating development strategies for a state, organization or country. The involvement of these initiatives in development is crucial. It may be for economic development, job-creation, rural development and poverty-alleviation etc. These initiatives have great potential to bring in the desired social transformation by enhancing its access to people, services, information and other technologies. Opportunities for the people can be enhanced by introducing IT applications by improving their access. Similarly citizens can be empowered by these initiatives through reaching out to them ensuring social and financial inclusion [2]. These initiatives can elevate living standards in remote and rural areas by providing important commercial, social and educational benefits [3].

We wanted to study the level of IT at grass root level and identify what constraints the monsoon of IT to enter rural India and flourish it. We started with identification of the village and then using interviews to assess the factors.

## II. SELECTION OF THE VILLAGE

The selection of village was done after a careful evaluation of many factors.

*1) Proximity to the technological hub:* The village should be near the capital Dehradun which is a thriving city of educational institutes; the hub of policy making and a very developed city of Uttarakhand. We wanted to see the conditions in a nearby village and then move towards other regions.

*2) The hilly state:* Uttarakhand being a hilly state the choice of our village should reflect the difficult geographical terrain as is the case with other villages of the state.

*3) Agro based: The village should be agro based Samalta satisfied all these conditions and hence was an ideal choice for the study.*

## III. SCOPE OF THE STUDY

*a) The study is carried out in Samalta Village*

*b) It uses the qualitative method of research*

## IV. OBJECTIVE

To identify the impact of Information technology in rural lives

## V. RESEARCH QUESTIONS

*1) What is the level of awareness with regards to IT?*

*2) What is the role of IT in Rural Areas?*

*3) Identification of Rural Needs for bottom up solutions through IT*

*4) Identifying possible visibility measures for rural resources.*

## VI. METHODOLOGY

### UNIT OF ANALYSIS

Individual residents of the village were interviewed.

### DESIGN OF THE STUDY

The village 'Samalta' is a mixed populated village with a population of about 1500. It is located about 75 K.m. away from Dehradun in the tehsil of Kalsi. In most of the cases, one or two male members from each family work outside the village for gainful employment. Agriculture and livestock are the traditional prime occupations. Contracting has been another common occupation usually carried out by the males.

Initially we visited the Gram Pradhan to know about the village and conduct a preliminary study.

Keeping the social, political, interpersonal and economic aspects in mind the interview method was followed where descriptive answers were received. With the dimensions identified by interviews with many experts, few factors were shortlisted to carry out the study in this remote village to measure the influence of the IT on village folk.

A principle of full disclosure was followed in order to make them feel comfortable and put them at ease to answer the questions. Each person spent about 15 minutes of their time to share their experience on their transition prior to and after IT initiatives. A personal narrative method of qualitative study was carried out with interview being semi structured and open ended.

The elements of the research were to:

a) *Identify the shared experience.*
b) *Explore the nature of the experience.*
c) *Examine the essence and the perspective of the phenomenon*

All the villagers shared their experience on the same phenomenon expressing their emotional, cognitive and gut feelings. The open-ended responses permit one to understand the world as seen by the respondents. We focused on use of open ended interviews to build up through personal narratives to determine if and how villagers see the role of IT. They were asked a series of questions about themselves, IT, their family and their household. In order to capture the actual words of the person being interviewed, their responses were recorded over the cell phone, Video camera and a few written notes. The recorded voice files enabled to give full attention to the respondents, build up eye contact and rapport and also be reflexive in terms of framing and reframing the questions in accordance with the responses. The villagers slowly shared their experiences, views and opinions and gradually revealed some of their innermost anguishes and aspirations in the course of the interviews.

Firstly, the recorded interviews which were in the local dialect of Hindi were transcribed and then translated into English and typed out. There were a few responses which were not translatable and hence the words and phrases were retained along with the translated version, to give a full flavour to the responses. The responses were grouped into categories best captured and to which they could fit into. The repetitive phrases or words in the responses were identified by the themes culled out by knowing what, which, where and how of the data. Textural description giving an idea of what they experienced and a structural description of how they experienced totally depicting the essence of the phenomena was identified.

## VII. FINDINGS

The important outcomes of the interviews are summarized in different headings

### INFORMATION ABOUT IT

We found that none of them could totality define IT the closest one dealt with information exchange. They were though of the view that IT is and could be beneficial.

"IT is all about information exchange"-Resp. 2

They had very little information about IT and its uses. They were of a notion that IT is all about transferring information from one person to another.

### LACK OF AWARENESS: THE BARRIER TO IT REVOLUTION

We found that people especially above 25 years are not aware about many IT tools and Govt. initiatives. This factor sprang up through many respondents

"Here people are not aware about IT" –Resp. 1

They admitted the lack of awareness among people about IT and related issues. Many referred to themselves and the other village folk when acknowledging the unawareness to IT initiatives and tools.

"We lack in awareness here"-Resp. 2

They were highly unaware about IT and its various applications.

"I don't have any knowledge about computer"-Resp. 4

They were of the view that there might be many tools and initiatives but also knew that the information would not reach them.

"People are not aware of the projects initiated by the Govt."-Resp. 9

They had no idea about the various projects and schemes launched by the Government. Also they were of the view that the schemes only benefit and reach to the urban population.

### ECONOMIC CONSTRAINTS: INHIBITOR OF IT WAVE:

Financing is a major problem to adopt IT. Even mobile is seen as convenient but expensive.

"Use of mobile can be a costly affair"-Resp. 7

They all had a common notion that usability of mobile is a costly affair. The comparative wealth of village is perceived to be lower than the urban setup.

"Financial status of the people is not up to the mark here"- Resp. 9

Computer is still perceived to be a costly machine and its usage in nearer to nil.

"Computer is a costly affair"-Resp. 8

"Villagers are not in a condition to afford computers"- Resp. 9

Financial status of the people is not up to the mark in Samalta. Therefore they were of the view that they cannot afford computer and its maintenance.

### DISFIGURED CONNECTIVITY: CRIPPLED LIFE OF IT

The connectivity wired and wireless is very poor and one of the major factor for stopping the free flow of information.

"There is a problem of internet connectivity here"-Resp. 2

"Connectivity is very poor here"-Resp. 9

They were of the view that connectivity is the major problem here and it hampers the free flow of communication.

Straddling a population of 740 million that logs in a GDP of over Rs. 600,000 cores, rural India presents enormous potential in thrusting India at the forefront of the most powerful nations of the 21st century. Connectivity is the key to harnessing the potential of its enormous human resources [4].

### GEOGRAPHICAL CONDITIONS: PREVENTING IT PENETRATION

We found that people of the village though being so close to the state capital find themselves isolated and deserted.

"We represent one of the remotest parts of Uttarakhand"- Resp. 1

A certain dissonance can be seen in people when they compare themselves to their urban counterparts, thus Geography has also induced this inferiority complex among villagers.

"We are still far behind when we compare ourselves to the metro cities and urban part of India" - Resp. 1

Because of the adverse geographical conditions, they are still forced to live in the absence of basic amenities. The pace of development has been slogging due to the difficult terrain and connectivity to this area.

"I think due to the geographical conditions of this region, the speed of IT revolution is very slow here" - Resp. 1

They also admitted that the geographical conditions of the area create an obstacle in the way to IT and related development.

India is a land of geographical diversities, WiMAX connectivity could play major role in improving the quality of public services and could bring substantial improvement in rural areas [5].

### RECEPTIVE AND OPTIMISTIC ABOUT IT: HOPES FOR CHANGE

They are of the view that IT has the capability to change their lives and thus help in personal and societal progress.

"People are more enthusiastic here"-Resp. 2

Even elders have an urge to learn and share the benefits of IT

"Yes why not, I would like to learn computers"-Resp. 3

They were passionate about learning computers. Elders showed the great interest in learning new technologies. They see IT education and benefits to be pervasive and for everyone.

"I think everybody should have knowledge of computers"- Resp. 5

They were of the view that the knowledge of computer is of utmost importance.

"Yes I would like to use computers"-Resp. 7

They are willing to undergo training to be at par with their urban counterparts.

"We would like to be a part of any training initiative"- Resp. 8

They really wanted to learn computers. They were of the view that training programmes must be conducted for them. They realise that IT has the potential to change and elevate their living standards.

"The use of tech will definitely help us; people are passionate about learning new things" -Resp. 9

"IT is very beneficial" - Resp. 1

They agreed upon the importance and advantages of IT. They also acknowledged the power of IT as a change agent.

### INITIAL FEAR OF IT: INERTIA FOR A CHANGE

The initial fear of using a new technology was evident in some cases.

"Before, I uses to be afraid of using mobile"-Resp. 4

The fear lowered after prolonged exposure to the IT device.

"I had to struggle a lot about its (computer) handling and I was scared at that time"-Resp. 5

Further research is recommended to provide a more holistic view of rural communities and their needs and of ways to develop ICT in these areas. Given the impact on attitudes to school and engagement with it as a result of the projects, research is also called for to explore how deep-seated antipathy to formal learning can be changed by community-based initiatives [6].

### IT THE TIME SAVER: QUICKNESS IS THE KEY

Mobile is seen as a portable source of communication and entertainment

"Yes, Mobile saves time"-Resp. 3

They agreed upon the view that mobile is a time saving device.

"Computer saves time and we can watch movies" Resp. 4

"Now we can talk to our relatives and dear ones through mobile and it saves lot of time"-Resp. 7

Those who use ATM see it as an easy and quick access to money.

"ATM saves lot of my time" -Resp. 9

"It helps us a lot .It saves our time it is beneficial in out day to day activities"-Resp. 5

Technology can help you save time, especially when you use the right technology and take the time to learn how to use it.

### IT AS BOOSTER OF THEIR PROFESSION:

People of the area are mostly farmers and even though they are not using IT they are hopeful that it will aid in their professional development.

"We can spread awareness about our products through computers"-Resp. 4

The knowledge of computer can prove to be an aid in their professions.

"It can be of great use in agricultural field also, in this hilly region most of us are dependent on agriculture as our livelihood" - Resp. 1

They believe that if they are connected to their customers they can eliminate middlemen who suck up a lot of chunk of the profit.

"Connectivity will add to our business prospects" -Resp. 9

They were of the view that proper connectivity will enhance their business prospects.

And it will also add to their economic status. They look forward to training from the Govt. to improve and re-engineer their work practices.

"Govt. should provide proper training to upgrade our business"-Resp. 3

They were of the view that Government should take initiative to conduct training programmes for their professional elevation and socio-economic development. The main profession being agriculture in village *IT for fields* could be the need of the grass roots.

"IT will create awareness about various innovative techniques in agricultural development" - Resp. 1

The application of Information and Communication Technology (ICT) in agriculture is increasingly important-

Agriculture is an emerging field focusing on the enhancement of agricultural and rural development through improved information and communication processes. More specifically, e-Agriculture involves the conceptualization, design, development, evaluation and application of innovative ways to use information and communication technologies (IT) in the rural domain, with a primary focus on agriculture. E-Agriculture is a relatively new term and we fully expect its scope to change and evolve as our understanding of the area grows.

The Veterinary Department of Malaysia's Ministry of Agriculture introduced a livestock-tracking program in 2009 to track the estimated 80,000 cattle's all across the country. Each cattle is tagged with the use of RFID technology for easier identification, providing access to relevant data such as: bearer's location, name of breeder, origin of livestock, sex, and dates of movement. This program is the first of its kind in Asia, and is expected to increase the competitiveness of Malaysian livestock industry in international markets by satisfying the regulatory requirements of importing countries like United States, Europe and Middle East. Tracking by RFID will also help producers meet the dietary standards by the halal market. The program will also provide improvements in controlling disease outbreaks in livestock [7].

### IT AS MIGRATION PREVENTER: STOPS THE OUTFLOW

They believe that if IT enables them to get visibility they would love to stay in the village instead of moving out to the cities.

"If we get better opportunities here then I don't think people will migrate to cities" - Resp. 1

The main reason of migration is lack of awareness and proper opportunities. They were of the view that people migrate in the absence of job prospects and better opportunities.

By 2030, India's urban population is set to reach 590 million, an addition of approximately 300 million to India's current urban population. Much of this growth will be due to rural-urban migration. The success of the Indian urbanization agenda will be hugely dependent on the poor migrants' integration as urban citizens [8].

### LACK OF RESOURCES: CURTAILING IT TO MASSES

The villagers are aware and disgruntled with the fact that lack of resources has kept them away from taking benefits of many IT tools and initiatives

"We lack in practical implementation of IT, we don't have sufficient resources here" - Resp. 1

Electricity, Mobile coverage, roads etc. are major resources which are needed for IT to flourish but these are in deficiency here.

"No I have never made my reservation done here, there is a problem of electricity here" - Resp. 1

"We had to struggle for even basic necessities like electricity and transport"-Resp. 4

They were of the notion that we had to even struggle for basic facilities like electricity, proper transportation facilities etc.

Despite several policy initiatives by the Government of India (GoI) and progress in extending the National grid, 56 % of rural households still do not have access to electricity. And even When they do, many have opted not to connect because of poor reliability and inadequate supply [9].

"Here we don't have good facilities"-Resp. 9

Resource, knowledge, status and technology are important factors for the development of Rural India. Rural folk have largely been ignorant to this fact as they were either confined to their immediate livelihood incomes they have been gaining through manual labour [10].

### GOVT.'S ROLE: WHAT SHOULD BE DONE?

Creating awareness and then usability could be seen as key challenges for Govt.; Good policy-making is only half of the solution. In the absence of proper execution or enforcement, it becomes mere eyewash, failing to help the most excluded.

"Govt. should help, motivate and make people aware here" - Resp. 1

Capacity building can be done through providing training to few who could impart it to others.

"There should a specialist to teach computers in  schools" - Resp. 1

An important aspect of maintainability of a measure taken by Govt. is highlighted.

"I think proper implementation of policies is very important, after framing policy it must be maintained properly" - Resp. 1

They were of the view that there should be proper implementation of the policies designed by the Government. Uniformity in implementation of policies will plug in the discrepancies

"Policies framed by the govt. must be implemented uniformly and properly. Facilities must reach villages"-Resp. 2

Subsidised training and IT equipment can be very beneficial.

"Govt. should do something for BPL families"-Resp. 3

"We should be equipped with computer"-Resp. 4

"Govt. must educate us about new trends in technological developments"-Resp. 5

"Govt. should frame good policies for the development of villages"-Resp. 6

"Yes…Govt. should train us"-Resp. 7

The need of the training is quite visible. They had ample of enthusiasm about learning new things. Also they were of the view that good training programmes must be conducted for them.

"We want proper connectivity here" -Resp. 9

### NEED FOR TRAINING: THE NEED OF THE HOUR

The need for training sprang up in many interviews; the bottom up demand from the grass roots can be cited as training and education.

"Yes…I see lots of benefits of training programmes and I think govt. must train villagers about new technologies and its use."-Resp. 2

"We should be trained in IT and its use"-Resp. 3

Training is linked to the level of awareness and thus prospect of use.

"Yes…why not, Govt. must focus on training people here as it leads to awareness" - Resp. 1

"Govt. must educate and train us towards this"-Resp. 5

"Yes the Govt. must train us, it would be beneficial for us"-Resp. 6

They see training as an important enabler for IT to touch their lives.

"We can't brush aside the importance of training"-Resp. 7

"Govt. should take initiative to start training program for us" -Resp. 9

### VIII.  CONCLUSION: PROPOSED GAPS MODEL

We identified many gaps between Govt and Public which are summarised below in the form of Gaps. A gap is evident only there is a mismatch between sending and receiving end. If there is a perfect match there will be no gap.

### The KDJ[1]-Gaps model

#### GAP 1: THE GOVT.-PUBLIC GAP

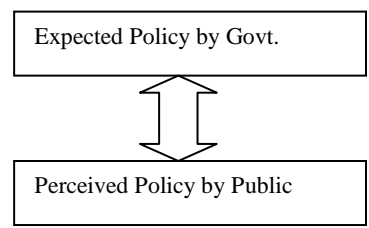

We find that there is a huge gap between the expected policy by Govt. and how it is perceived by Public. Govt. expects a policy like Aadhar for social inclusion of the deprived but we found many of them see it as just substitute to Ration card.

---

1 KDJ refers to the authors' initials

### GAP 2: THE PUBLIC GOVT. GAP

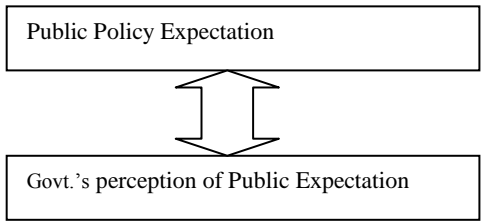

Public expects policies to be tailored to their needs but Govt. due to lack of research perceives the need differently and designs mismatching policies which don't benefit public at large.eg. Computers were provided at the village Samalta but no training on computers was provided either to villagers or teachers.

### GAP 3: THINKING ACTION GAP

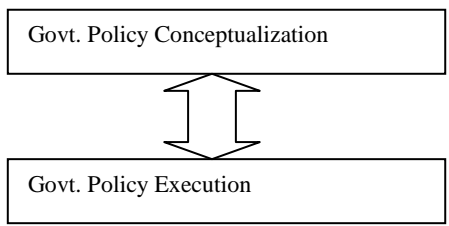

The conceptualization of a policy may be best but the execution makes it worst. The execution kills the essence of a policy.

### GAP 4: MAINTAINABILITY GAP

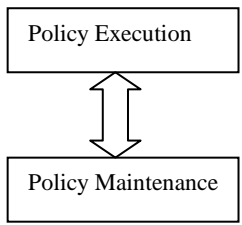

The poor maintenance of the policy eventually leads to non-execution of the policy. We found in our study that the computers for the village school were dumped because they were not being maintained.

### GAP 5: COMMUNICATION GAP

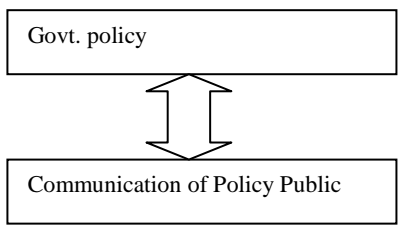

The distortion in communication or sometimes no communication has been major cause of low awareness towards IT initiatives.

We can conclude that the Govt.'s pull of policies towards public is not equal to the demand or pull by public and hence the cause of discrepancies.

Govt Push   !=Public  Pull

### ACKNOWLEDGEMENT

We thank all the villagers of Samalta Village for their support during the course of the study.

### AUTHORS PROFILE

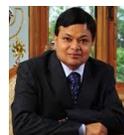
Prof. K.K. Ghanshala is the chairman of Graphic Era University, Dehardun, the capital of Uttarakhand, India. He founded Graphic Era in the year 1993 as one of his bold ventures towards providing quality educational inclusion to the society at large and people living in difficult geographies in particular.

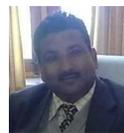
Dr. Durgesh Pant is Professor of Computer Science, Department of computer science at Kumaun University, Nainital, Uttarakhand which he founded way back in 1989. Presently, he is working as Professor and Director of the School of Computer Science & IT at Uttarakhand Open University, Campus Dehradun, India. To Prof. Pant's credit goes the distinction of taking computing & informatics to its present level in this part of the world.

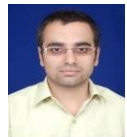
Jatin Pandey is working as J.R.F. in Graphic Era University, Dehradun. His research areas are multidisciplinary research in Management, Computer Science and Total Quality Management.